\documentclass[journal,final,10pt]{IEEEtran}
\usepackage[ruled,vlined]{algorithm2e}
\usepackage{algorithmic}
\usepackage{hyperref}
\hypersetup{colorlinks,
citecolor=blue, 
filecolor=black,
linkcolor=black,
urlcolor=black}
\usepackage{amsthm}
\usepackage{stfloats}
\usepackage{blindtext}
\usepackage{multicol,lipsum}

\usepackage{times}
\usepackage[cmyk]{xcolor}
\usepackage{color} 

\usepackage{bm,mathptmx,amsfonts,amssymb,amstext,textcomp}
\usepackage{epsfig,graphicx,url,graphics,epstopdf,multirow}
\usepackage[cmex10]{amsmath}
\usepackage{mathtools}
\usepackage{cuted}

\newcommand{\rea}{\mathbb{R}}

\newtheorem{mythm}{Theorem}
\newtheorem{mypf}{ Proof of Theorem}
\newtheorem{mypflem}{\bf Proof of Lemma}
\newtheorem{myass}{\bf Assumption}
\newtheorem{mylem}{\bf Lemma}
\newtheorem{myrem}{\bf Remark}

\newtheorem*{spm*}{\bf Singular Perturbation Method}

\DeclareMathOperator{\sign}{sign}

\DeclareMathOperator{\G}{\mathcal{G}}

\DeclareMathOperator{\V}{\mathcal{V}}
\DeclareMathOperator{\E}{\mathcal{E}}

\DeclareMathAlphabet\mathcal{OMS}{cmsy}{b}{n}

\ifCLASSINFOpdf
\else
\fi

\begin{document}

\title{Leader-Following Consensus of High-Order Perturbed
	Multi-agent Systems under Multiple Time-Varying  Delays}

\author{Milad~Gholami
\thanks{This work is partly financed by the Italian
	Ministry for Research in the framework of
	the 2017 Program for Research Projects of
	National Interest (PRIN) under Grant
	2017YKXYXJ.}
\thanks{Milad Gholami is Department of Electrical
	and Mathematics, University of Siena,
	Siena, Italy. Email: gholami@diism.unisi.it.} }

\maketitle
\begin{abstract}
Solving an output consensus problem in multi-agent systems is often hindered by multiple time-variant delays. To address such fundamental problems over time, we present a new optimal time-variant distributed control for linearly perturbed multi-agent systems by involving an integral sliding mode controller and a linear consensus scheme with constant wights under directed topology. Lyapunov-Krasovskii functionals along with linear matrix inequalities are jointly employed to demonstrate the associated closed-loop stability and convergence features. Maximum delays for the communicating networks are also estimated by linear matrix inequalities. Synchronizing a network of linear time-variant systems to the associated leader dynamics is additionally taken into account by developing an optimization algorithm to find the constant control gains.
\end{abstract}

%
\IEEEpeerreviewmaketitle
\section{Introduction}
%
%
%
%
\IEEEPARstart{M}uLTI-AGENT Systems (MASs) are extensively networked to describe very large numbers in natural and engineered environments, ranging from biological, social and communication systems \cite {8740876,8585045,7403961} to transport, power grids and robotic swarms \cite{gholami2018robust,8643771,gholami2021robust}. The problem with their consensus, however, remains an attractive challenge in controlling multiple interacting agents within the system. Consensus can thereof be regarded as a control objective in which all the agents in a network converge to (or, agree upon) a common value. This is achieved through a given control strategy, referred to usually as a consensus algorithm. Most of the early works on this topic \cite{olfati2007consensus,1333204,7438822} were based on the leaderless consensus in multi-agent systems, whereas all participating agents had to agree with certain coordinate values as problems. Similar studies were then shifted toward a leader-follower approach, wherein a network of follower-agents has to be regulated by the leader coordinates \cite{7891050,8290709,8726137,gholami2019distributed}. 

Inclusion of time-variant delays in any multi-agent model has become a challenge due to the limitation of axonal signal transmission and switching speeds. Being inevitable in many physical systems, a time delay, as an integral part of the convergence in consensus protocols, is therefore to be studied \cite{wu2019class}. The stability of a single-agent system with time delays are realized 
\cite{rosinberg2018influence,klinshov2017embedding,7432018}, with emphasis on the effect of a delay in exchanging the information in linear stochastic models \cite{rosinberg2018influence}. The connection between the dynamics of a single oscillator with delayed feedback and a feedforward ring of identical oscillators was also highlighted \cite{klinshov2017embedding}. Moreover, the real-time data was recorded by a delay scheduled impulsive controller \cite{7432018}. In this context, a multi-agent model with input and constant communication delays can then be analysed in frequency-domain to underline the fact that the consensus is indeed independent of the communication delays \cite{tian2008consensus}. A distributed and robust $H\infty$ rotating consensus control within a single time delay was then reported \cite{ping2013distributed}. In addition, the problems of synchronization and nonmonotonic transitions in oscillator communities with distributed constant delays were investigated \cite{restrepo2019competitive}. As for synchronizing a heterogeneous network, multiple constant time delays were similarly involved \cite{otto2018synchronization}. Consensus problems of MASs within second-order continuous time were as well investigated by considering single time delays and jointly-connected topologies \cite{lin2010consensus}.
 
The cases for continuous-time and discrete-time first-order MASs with uniform input constant delays could marginalize the maximum consensus delay \cite{xu2013input}, whereas general second-order models with single constant delay conditioned by an integrator consensus problem \cite{hou2017consensus}. Instantaneous states of the leader itself and the delayed states of its followers were also focused upon   \cite{8721632}. In result, a distributed control algorithm with multiple constant delays for these second-order models was proposed \cite{li2017further}. More to this, the delayed state synchronization of homogeneous discrete-time MASs in the presence of unknown non-uniform communication delays was studied \cite{liu2018passivity}. An observer-based triggering control problem was then based for leader-following consensus of MASs with time-varying delays \cite{wang2017observer}. Another consensus problem of discrete-time linear MASs was then controlled by a distributed prediction within directed switching topologies and constant delays \cite{7961224}. Nonlinear agents, however, were synchronized in directed networks within single delays \cite{8782631}. 

It is worth to mention that most of these consensus problems are bounded to only a single delay, particularly in those of systems where the agent dynamics, either single or double, integrates with specific consensus protocols. We herein introduce a new linear distributed control in MASs within multiple time-varying delays. This is achieved by combining an integral sliding mode controller with tunable constant weights in well-networked perturbed linear MASs. Compared to that of the existing models, the performance of the proposed scheme is thoroughly analyzed by combining both the Lyapunov- Krasovskii theorem and the Linear Matrix Inequality (LMI) approach. An upper bound for maximum tolerable input delays through LMIs is also provided. Furthermore, the net gain of the proposed control is tuned and optimized.

As follows, preliminaries in the section~\ref{sec2} present the basic concepts of algebraic graph theory. Section~\ref{sec3} introduces the problem statement. Our main results are then discussed in the Sections~\ref{sect4} and \ref{sect5}. Section~\ref{sect6} details the numerical models, and finally, the conclusions are collected in the section~\ref{sect7}.

\section{Preliminaries} \label{sec2}
The set of natural, real, and strictly positive real numbers are denoted by $\mathbb{N}$, $\mathbb{R}$ and $\mathbb{R}^{> 0}$, respectively. For $d \in \mathbb{N}$ and  a column vector of $x \in \mathbb{R}^d$, let $x'$ be its transpose and $\Vert x\Vert_2$ the corresponding 2-norm. Then, define $\mathcal{G}=\left(\mathcal V,\mathcal E\right)$ as a directed graph (digraph), where $\mathcal V=\lbrace 1,\ldots,N\rbrace$ is a set of nodes (i.e. agents) and $\mathcal E\subseteq \{\V\times \mathcal V\}$ is the set of edges. $\mathcal A = [\alpha_{ij}]\in \mathbb{R}^{N \times N} $ is the adjacency matrix of $\G$, with non-zero weight if $i$ communicates with $j$ $(i; j)\subseteq\E$, $\alpha_{ij}= 0$ otherwise. Let ${\mathcal N}_i=\{j \in \V:(i,j)\in \E\}$ be the set of neighbors around agent $i$, showing those of agents that share an edge with agent $i$. Next, let agent “0” be an additional virtual object in the augmented
graph $\G_{N+1}$, while the remainder “0” is considered as the virtual leader for the proposed protocol assuming to be globally reachable in $\G_{N+1}$.

In the remainder of the paper the following useful properties are exploited:
\begin{mythm}
	Suppose that $f:\mathbb{R} \times C \in ({\small\begin{bmatrix}
		-h&0 \\
		\end{bmatrix}},\mathbb{R}^n)\longrightarrow \mathbb{R}^n $ in $\dot{x}(t)=f(t,x_{t})$ maps $\mathbb{R}\times$ (bounded sets of $C$) into bounded sets of $\mathbb{R}^n$ and also that $u, v, w :$ ${\mathbb{R}}_{>0}\longrightarrow {\mathbb{R}}_{>0}$ are continuous non-decreasing functions, $u(s)$ and   $v(s)$ are positive for $s>0$, $u(0)=v(0)=0$ and $v$ is always increasing. If there exists a continuously differentiable function $V:\mathbb{R} \times \mathbb{R}^n\longrightarrow \mathbb{R} $ such that
	\begin{equation}
	u(\Vert x \Vert)\leq V(t,x(t))\leq v(\Vert x \Vert),\quad t\in\mathbb{R}, \quad x\in\mathbb{R}^n,
	\end{equation}
	and the derivative of $V(t,x(t))$ along the solution, $x(t)$, of $\dot{x}(t)=f(t,x_{t})$ satisfies
	\begin{equation}
	V(t,x(t))\leq - w(\Vert x \Vert)
	\end{equation}
	whenever
	\begin{equation} \quad \quad V(t+\theta,x(t+\theta))\leq V(t,x(t))
	\end{equation}
	for $\theta \in {\small\begin{bmatrix}
		-h&0 \\
		\end{bmatrix}}$ where $h>0$ is the delay, then the trivial solution of $\dot{x}(t)=f(t,x_{t})$ is uniformly stable.{\hfill $\blacksquare$}
\end{mythm}
\begin{mylem}\label{lem1}
	The Schur complement lemma converts a class of
	convex nonlinear inequalities that appears regularly in
	control problems to an  LMI. The convex nonlinear
	inequalities are
	\begin{equation}
	R(x)<0,\quad \quad 		Q(x)-S(x)R(x)^{-1} S(x)'<0,
	\end{equation}
	where $Q(x)= Q(x)'$, $R(x)= R(x)'$, and $S(x)$ depend
	affinely on $ x$. The Schur complement lemma converts
	this set of convex nonlinear inequalities into the
	equivalent LMI
	\begin{equation}
	{\small\begin{bmatrix}
		Q(x) & S(x) \\
		S(x)' & R(x)
		\end{bmatrix}}<0
	\end{equation}{\hfill $\blacksquare$}
\end{mylem}
\begin{mylem}\label{lem2}
	By letting $a(t),b(t) \in 	\rea^{n}$ and $\psi \in \rea^{n\times n}_{>0}$, the following inequalities are in force
	
	\begin{flalign}
	\vert 2a(t)'b(t)\vert&\leq a(t)'\psi a(t)+b(t)'\psi^{-1}b(t).\label{Eeq:1lem2}\end{flalign}
	Moreover, let $h\in\rea$ and $\bar{h}$ be the maximum value assumed by a time delay, then
	\begin{equation}
	-2a'\int_{t-h(t)}^{t}\dot{b}(s)ds\leq \bar h a(t)'\psi^{-1}a+\int_{t-h(t)}^{t}\dot{b}(s)\psi \dot{b}(s)ds\label{Eeq:2lem2}
	\end{equation}{\hfill $\blacksquare$}
\end{mylem}
\begin{mythm}\label{thm1}
	Let $M_1 \in \mathbb{R}^{n \times n}$ be a negative symmetric matrix and $M_2\in \mathbb{R}^{n \times n}$  be a positive definite matrix, and then let $\tau\in\rea_{> 0}$ be a positive constant. If the following relation is in force:
	$$M_1+\tau M_2<0$$
	then it yields
	$$M_1+\xi M_2<0 \quad \forall~\xi ~(0,\xi_m] $$
	where
	$$\xi_m=\frac{\Vert M_1 \Vert}{\Vert M_2 \Vert} $${\hfill $\blacksquare$}
\end{mythm}
\section{Problem Statement}
\label{sec3}
We consider a continues-time MAS with topology represented
by a directed graph $\mathcal{G}=\left(\mathcal V,\mathcal E\right)$ where the agents have the
dynamics described by
\begin{flalign}
\dot{x}_i(t)&= \displaystyle Ax_i(t)+B u_i(t)+B\omega_i(t)\label{eqn:1}
\end{flalign}

where $x_i\in\mathbb{R}^{n}$ represents the state of the $i$-th agent and $u_i(t)\in \mathbb{R}$  is
a control protocol which needs to be designed.{ The partial function of $\omega_i(t) \in \mathbb{R}$, denotes unknown exogenous perturbations.} And also $A \in \mathbb{R}^{n \times n}$ and $B\in\mathbb{R}^{n}$ have the following expression:
\begin{flalign}A={\small\begin{bmatrix}
	0 & 1 & 0 &\ldots & 0 \\
	0 & 0 & 1 &\ddots & 0 \\
	\ldots & \ldots & \ldots &\ddots &\vdots\\
	0 & 0 & 0 &\ddots &1\\
	-a_1 & -a_2 & \ldots &\ldots &-a_n
	\end{bmatrix}}\quad B={\small\begin{bmatrix}
	0 \\
	\vdots \\
	0\\
	b
	\end{bmatrix}}\label{eqn:nn1}\end{flalign}
with $b > 0$ and $a_1,a_2,\ldots,a_n$ are positive constants.
\begin{myass}\label{ass1}
{ We assume that for the system \eqref{eqn:1}, the
unknown perturbations are bounded:
\begin{flalign}
|\omega_i(t)|&\leq \Gamma_i\leq \Gamma\quad \Gamma=\mathrm{\max_{i\in{\mathcal{V}}}}\lbrace\Gamma_i\rbrace \leq \infty,\quad \forall~i\in \mathcal{V}\label{eqass1}\end{flalign}
where $\Gamma_i$ is a positive known constant}.{\hfill $\blacksquare$}
\end{myass}
Our objective is to design a continues-time distributed consensus
protocol $u_i(t)$ for agents as in Eq.~\eqref{eqn:1} to enable each agent to track
the time-varying virtual leader despite delayed communications among agents, wherein the reference dynamic is described as 
\begin{flalign}
\dot x_{0}&=Ax_0(t)\label{eqn:2}
\end{flalign}
\section{Proposed Distributed Consensus Protocol}\label{sect4}
To synchronize the agent's states to the state's virtual leader, we present the following local interaction protocol of
\begin{align}
\tiny
u_i(t)=&-\sum_{j=0}^{N}\alpha_{ij} k_{ij}
( x_{i}(t-\tau_{ij}(t))-x_{j}(t-\tau_{ij}(t)))\nonumber\\
&-\rho\sign\big(s_i(t)\big) \label{eqn:1nn}
\end{align}
{ where $k_{ij}' \in \mathbb{R}^{n}$ and $\rho\in\mathbb{R}$ are the constant tuning-parameter vectors and scalars, respectively}. $\tau_{ij}(t)$ shows the time-varying delay between the communicating agents. We also assume that
$\tau_{ij}(t)=\tau_{ji}(t)$. {The
switching function $s_i(t)$ in \eqref{eqn:1nn} set as follows:}
\begin{equation}
s_i(t)=  \bold{1}_n'x_i(t)+z_i(t)\label{slide}
\end{equation}
where
\begin{gather}
\dot z_i(t)=  \bold{1}_n'\Big(Ax_i(t)+B\sum_{j=0}^{N}\alpha_{ij} k_{ij}
\big( x_{i}(t-\tau_{ij}(t))-x_{j}(t-\tau_{ij}(t))\big)
\Big)\nonumber\\
z_i(0)=-\bold{1}_n'x_i(0)
\label{eqnnqr}
\end{gather}
To exploit a more compact notation, delays $\tau_{ij}(t)$ can be represented as elements of the following delay set:
$
\sigma_ p(t)\in  \left\lbrace \tau_{ij}(t):i,j=1,\ldots,N,i\neq j\right\rbrace
$ for $p=1,\ldots,m$ with $m\leq N(N-1).$ Analogously, delays $\tau_{i0}$ are elements of the set:
$
\tau_l(t)\in  \left\lbrace \tau_{i0}(t):i=1,\ldots,N\right\rbrace
$
for $l=1,\ldots,N.$

The next assumption refers to \cite{liu2017robust}-\cite{ko2018effect} to solve the communication's delays.
\begin{myass}\label{ass2}
Let the known bounds to $\tau_l^{\star}$,$\sigma_p^{\star}$,$d_l$ and $\bar d_p \in\rea_{> 0}$ exist and be known in advance such that
\begin{flalign}	\tau_{i0}(t) \in&  \left[0,\tau_l^{\star}\right)\quad \forall ~ t\geq0,~\forall~\tau_{i0}\in\tau_l,~l=1,2,\ldots,N.\nonumber\\	\tau_{ij}(t) \in&  \left[0, \sigma_p^{\star}\right)\quad\forall ~ t\geq0,~\forall~ \tau_{ij}\in\sigma_p,~p=1,2,\ldots,m.\nonumber\\|\dot\tau_{i0}(t)|&\leq d_l<1\label{eqn:MY5}\\	|\dot\tau_{ij}(t)|&\leq\overline{d}_p<1\nonumber	\end{flalign}{\hfill $\blacksquare$}
\end{myass}
\begin{mythm}\label{thm2}\it Consider the multi-agent system dynamics \eqref{eqn:1} operating over a communication network whose topology can be described by a directed connected graph $\G$. Let Assumption~\ref{ass2}  be satisfied and there exits $\rho>\varGamma$. Then let \eqref{eqn:1nn} be local control protocol for the $i$-th agent communication and assume the state of the agent “0” be globally reachable over $\G_{N+1}$. Given an upper bound of time-delay function $\tau^{\star}=max\left\lbrace \tau_l^{\star},\sigma_p^{\star}\right\rbrace >0$, If there exist symmetric positive definite matrices $P, Q_l, \bar Q_p, R_l$ and $\bar R_p\in \mathbb{R}^{Nn\times Nn}$, such that the following LMIs are feasible
	\begin{equation}{\small
		{\small\begin{bmatrix}
			M_1+3\tau^{\star}A_0'HA_0 & \sum_{l=1}^{N}\tau^{\star}P\hat A_l&\sum_{p=1}^{m}\tau^{\star} P\tilde A_p  \\
			* & -\sum_{l=1}^{N}\tau^{\star}R_l&0_{Nn \times Nn}\\
			* & *&-\sum_{p=1}^{m}\tau^{\star}\bar R_p
			\end{bmatrix}}< 0 } \label{eqn:49}
	\end{equation}
	\begin{equation}
	3\sum_{l=1}^{N}\tau^{\star}\hat A_l'H\sum_{l=1}^{N}\hat A_l+\bar M_1< 0
	\end{equation}
	\begin{equation}
	3\sum_{p=1}^{m}\tau^{\star}\tilde A_p'H\sum_{p=1}^{m}\tilde A_p+\bar M_1< 0 \label{eqn:50}
	\end{equation}
being
	\begin{flalign}
	H&=\sum_{p=1}^{m}\bar R_p+\sum_{l=1}^{N}R_l\label{H}\\
	M_1&=F'P+PF+\sum_{p=1}^{m}\bar Q_p+\sum_{l=1}^{N}Q_l\label{m1}\\
	\bar M_1&=\sum_{l=1}^{N}-Q_l(1-d_l)\label{m1h}\\
	\tilde M_1&=\sum_{p=1}^{m}-\bar Q_p(1-\bar d_p)\label{m1T} \end{flalign}
	\begin{flalign}
	A_{0}&=\mathrm{diag}\left([A,\dots,A]\right)\in \mathbb{R}^{Nn\times{ Nn}}\label{eqn:n51}
	\end{flalign}
	\begin{equation}
	{\small \hat A_l=\mathrm{diag}\left([A_{(1,1)},\dots,A_{(N,N)}]\right)\in \mathbb{R}^{Nn\times{ Nn}}}\label{eqn:52}
	\end{equation}
	with diagonal blocks as
	\begin{equation}{\small
		A_{(i,i)}=\begin{cases}
		A_{i0} &i =l,~\tau_l(\cdot)=\tau_{i0}\\
		0_{n\times n}& i\neq l,~\tau_l(\cdot)\neq\tau_{i0}.
		\end{cases}}
	\label{eqn:mn53}
	\end{equation}
	and matrix $\tilde{A}_p=[\tilde{A}_{p(r,q)}]\in \mathbb{R}^{Nn\times Nn}$ within the following entries
	\begin{equation}{\small
		\tilde{A}_{p(r,q)}=\begin{cases}
		A_{ij} & \mbox{with }i\neq j\quad \mbox{if~}\sigma_p(\cdot)=\tau_{ij}(\cdot),r=q=i\\
		-{A}_{ij}& \mbox{with} ~i\neq j\quad \mbox{if}~\sigma_p(\cdot)=\tau_{ij}(\cdot),r=i,q=j \\
		0 & \mbox{otherwise}.
		\end{cases}}
	\label{eqn:n53}
	\end{equation}
	with \begin{align}
	{A}_{i0}&=-B\alpha_{i0}\cdot {k}_{i0} \in\mathbb{R}^{n \times n}\label{eqn:66}\\
	{A}_{ij}&=-B\alpha_{ij}\cdot k_{ij}\in\mathbb{R}^{n \times n}\label{eqn:n66}
	\end{align}
	where $r,q=\left\lbrace 1,2,\ldots, N\right\rbrace $, and
	\begin{equation}{\small
		F=A_{0}+\sum_{l=1}^{N}\hat{A}_l+\sum_{p=1}^{m}\tilde{A}_p}\label{eqn:73}
	\end{equation}
	Then, the delayed MAS \eqref{eqn:1} under constant gains $k_{ij}$, achieves synchronization,
	i.e.
	\begin{equation}
	\Vert x_i-x_0 \Vert\longrightarrow 0\quad \forall i \quad as~t\longrightarrow \infty
	\label{eqn:syn}\end{equation}{\hfill $\blacksquare$}
\end{mythm}
\begin{myrem}In accordance with Theorem~\ref{thm1}, it results that agents \eqref{eqn:1} under the control protocol \eqref{eqn:1nn} perform synchronization on the leader's states in accordance with \eqref{eqn:syn}. From \eqref{eqn:88}-\eqref{eqn:89} it further results that an estimation of the maximum admissible delay tolerated by MASs \eqref{eqn:1}-\eqref{eqn:1nn}, where it can be lower estimated as follows
\begin{flalign}
\tau^{\star}=\min\left\lbrace \frac{\Vert M_1\Vert}{\Vert M_2\Vert},\frac{\Vert \bar M_1\Vert}{\Vert \bar M_2\Vert},\frac{\Vert \tilde M_1\Vert}{\Vert \tilde M_2\Vert} \right\rbrace \label{eq:mar}
\end{flalign}{\hfill $\blacksquare$}
\end{myrem}
\begin{myrem}
{ Due to the discontinuous nature of \eqref{eqn:1nn}, high-frequency chattering on the agents state variables arises in practical implementation. To relax this phenomenon, several useful methods were proposed in \cite{lee2007chattering},\cite{utkin1996integral}. We thereby reformulate \eqref{eqn:1nn} as: 
\begin{align}
\tiny
	&u_i(t)=-\sum_{j=0}^{N}\alpha_{ij} k_{ij}
	( x_{i}(t-\tau_{ij}(t))-x_{j}(t-\tau_{ij}(t)))-\rho\upsilon_i(t)\nonumber \\
	&\widetilde{T}\cdot\dot{\upsilon}_i(t)+{\upsilon}_i(t)=\rho\sign\big(s_i(t)\big) \label{chatring2}
	\end{align}
by which a relatively high gain filtering is retained to validate an equivalent control, and subsequently estimate the perturbation in the system by chattering alleviation method. Note that if the time constant
 $\widetilde{T}$ of the filter is small enough such that the filter preserves the slow component of an equivalent control, and under the realistic assumption that the spectrum of the perturbation $\omega_i(t)$ does not overlap with the high-frequency components of the switching control, then:
	$$\upsilon_i(t)\approx\sign(s_i(t))=\frac{1}{\rho} \omega_i(t)$$
so that the unwanted chattering effect is mitigated, while the accuracy of the original discontinuous system is better preserved compared to more of the conventional saturation-based chattering alleviation methods.}{\hfill $\blacksquare$}
\end{myrem}	
\section{Solving optimization over LMI} \label{sect5}
In this section, we explain how to solve the LMIs feasibility problem \eqref{eqn:49}-\eqref{eqn:50} so that  we can find the robust values of the constant gains $k_{ij}$ that guarantee the output consensus of the MAS in \eqref{eqn:1} to the leader dynamics in \eqref{eqn:2}. This problem can be converted into the following LMI
optimization
\begin{flalign}
\max_{k_{i0,k}, k_{ij,k}}&\min\left\lbrace \frac{\Vert M_1\Vert}{\Vert M_2\Vert},\frac{\Vert \bar M_1\Vert}{\Vert \bar M_2\Vert},\frac{\Vert \tilde M_1\Vert}{\Vert \tilde M_2\Vert}  \right\rbrace \label{eqn:nn54}
\end{flalign}
\begin{equation}{\small
	S.t~\begin{cases}
	k_{ij,k}>0\quad(k=1,\ldots,n.\quad i,j=1,\ldots,N)\\
	\phi_i<0
	\end{cases}}  \label{eqn:nn53}
\end{equation}

\begin{figure}[htp]
	\centering
	\includegraphics[width=1\linewidth]{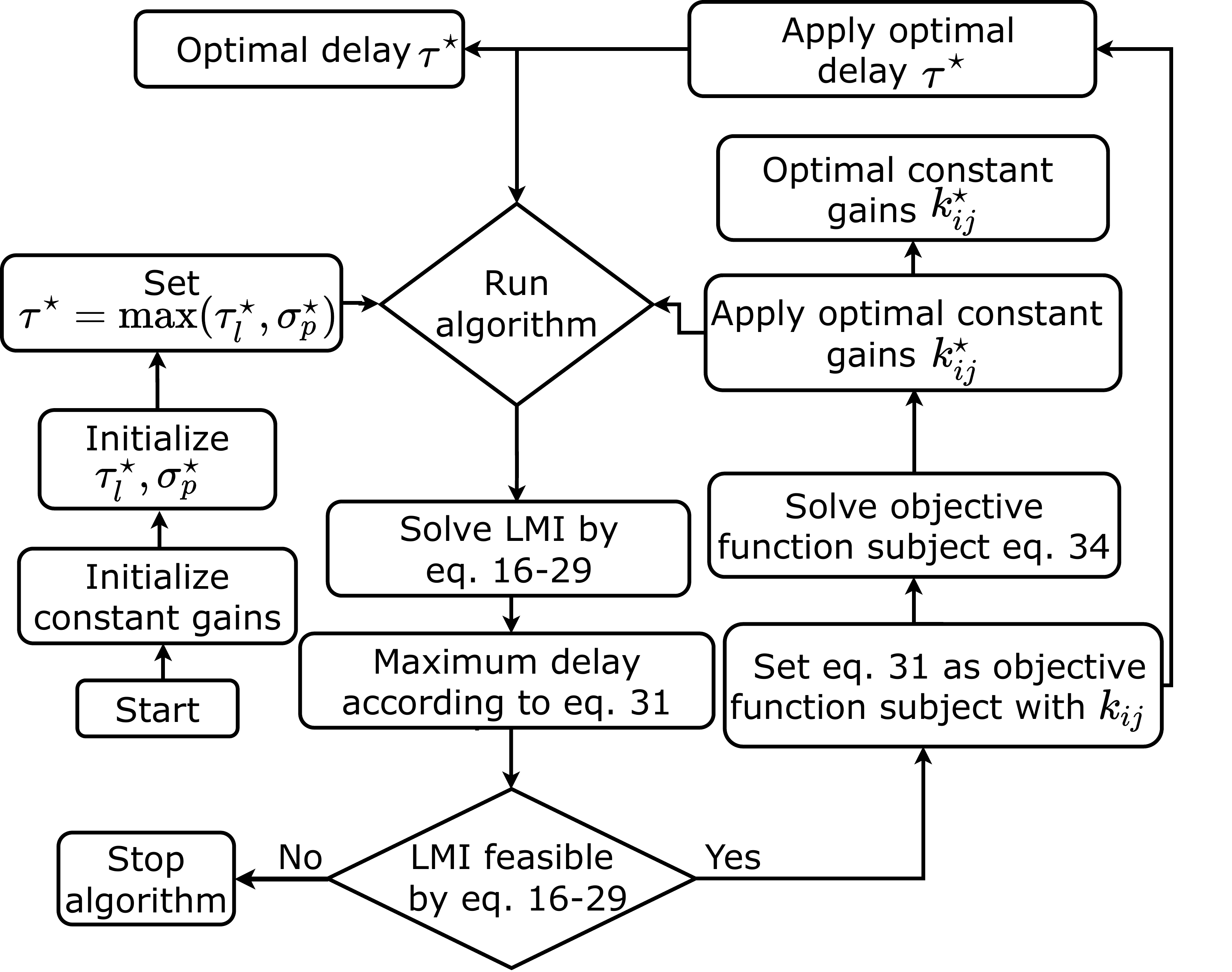}
	\caption {Process flow diagram of the developed optimization algorithm.}
	\label{figurelabel1}
\end{figure}

In the proposed optimization, we obtain the largest upper bound of the delay  by solving \eqref{eqn:nn54}-\eqref{eqn:nn53} within the
variables $k_{ij}$. The process of using the optimization algorithm to solve the
LMI is shown in Fig.~\ref{figurelabel1} and summarizes in Algorithm~1.
According to this optimization algorithm, we find the maximum input delays and the best tuning gains $k_{ij}$ such that the system \eqref{eqn:1} remains consensusable under the proposed
protocol \eqref{eqn:1nn}.
\begin{algorithm}[htp]
	\SetAlgoLined
	\vspace{1mm}
	$\bullet$ Initialize constant gains $k_{ij}$.\\ 
	$\bullet$ Given upper bound of time delays as $\tau_l^{\star}$ and $\sigma_g^{\star}$.\\
	$\bullet$ Consider $\tau^{\star}=max\left\lbrace \tau_l^{\star},\sigma_g^{\star}\right\rbrace $.\\ 
	\Repeat{LMIs problem \eqref{eqn:49}-\eqref{eqn:73} be feasible.}{
		$\bullet$ Solve the LMIs feasibility problem \eqref{eqn:49}-\eqref{eqn:73} with $\tau^{\star}$.\\
		$\bullet$ Estimation of the maximum admissible delays according to \eqref{eq:mar}.\\
		$\bullet$ Check the LMIs feasibility problem \eqref{eqn:49}-\eqref{eqn:73} according to the obtained maximum delay.\\ 
		\eIf{LMIs problem \eqref{eqn:49}-\eqref{eqn:73} are not feasible}{$\bullet$ Break.
		}
		{
			$\bullet$ Solve objective function \eqref{eqn:nn54} subject to \eqref{eqn:nn53}.\\
			$\bullet$ Apply optimal constant gains $k_{ij}^{\star}$ and optimal delay $\tau^{\star}$ to LMI problems.\\
		}
		$\bullet$ Return optimal constant gains $k_{ij}^{\star}$.\\
		$\bullet$ Return maximum delay $\tau^{\star}$.\\
	}
	\caption{Optimization algorithm}\label{tab1}
\end{algorithm}
\section{Verification of Results}\label{sect6}
To test the performance of the proposed protocol \eqref{eqn:1nn}, we consider here a generic MAS composed by 4 agents plus a leader. The $i$-th agent dynamics are defined as
\begin{flalign}\dot x_i=\begin{bmatrix}
0 & 1& 0  \\
0 & 0& 1  \\
-1 & -2&-3
\end{bmatrix}x_i(t)+\begin{bmatrix}
0 \\
0 \\
1
\end{bmatrix}u_i(t)\label{eqn:nnn1}\end{flalign}
and the leader dynamic is considered as
\begin{flalign}\dot
x_0=\begin{bmatrix}
0 & 1& 0  \\
0 & 0& 1  \\
-1 & -2&-3
\end{bmatrix}x_0(t)\label{eqn:nnn2}\end{flalign}

According to Routh–Hurwitz stability, $\phi_i$ in \eqref{phi} is negative if the following condition satisfies
\begin{equation}
\gamma_3\cdot\gamma_2>\gamma_1\label{bn}
\end{equation}
From \eqref{phi1} and \eqref{eqn:nnn1}, \eqref{bn} can be recast as
\begin{equation}
k_{i0,1}-2k_{i0,2}-3k_{i0,3}-k_{i0,2}\cdot k_{i0,3}<0
\end{equation}
The communication topology $\mathcal{G}$ is chosen in accordance to the following
adjacency matrix
\begin{flalign}
\mathcal{A}=\left[\begin{matrix}
0 & 0 & 1 & 0 \\
0 & 0 & 0 & 1 \\
0 & 0 & 0 & 1 \\
1 & 1 & 0 & 0
\end{matrix}\right]
\label{eqn:30}\end{flalign}
\newpage
\begin{figure}[htp]
	\centering
	\includegraphics[width=1\linewidth,height=0.65\linewidth]{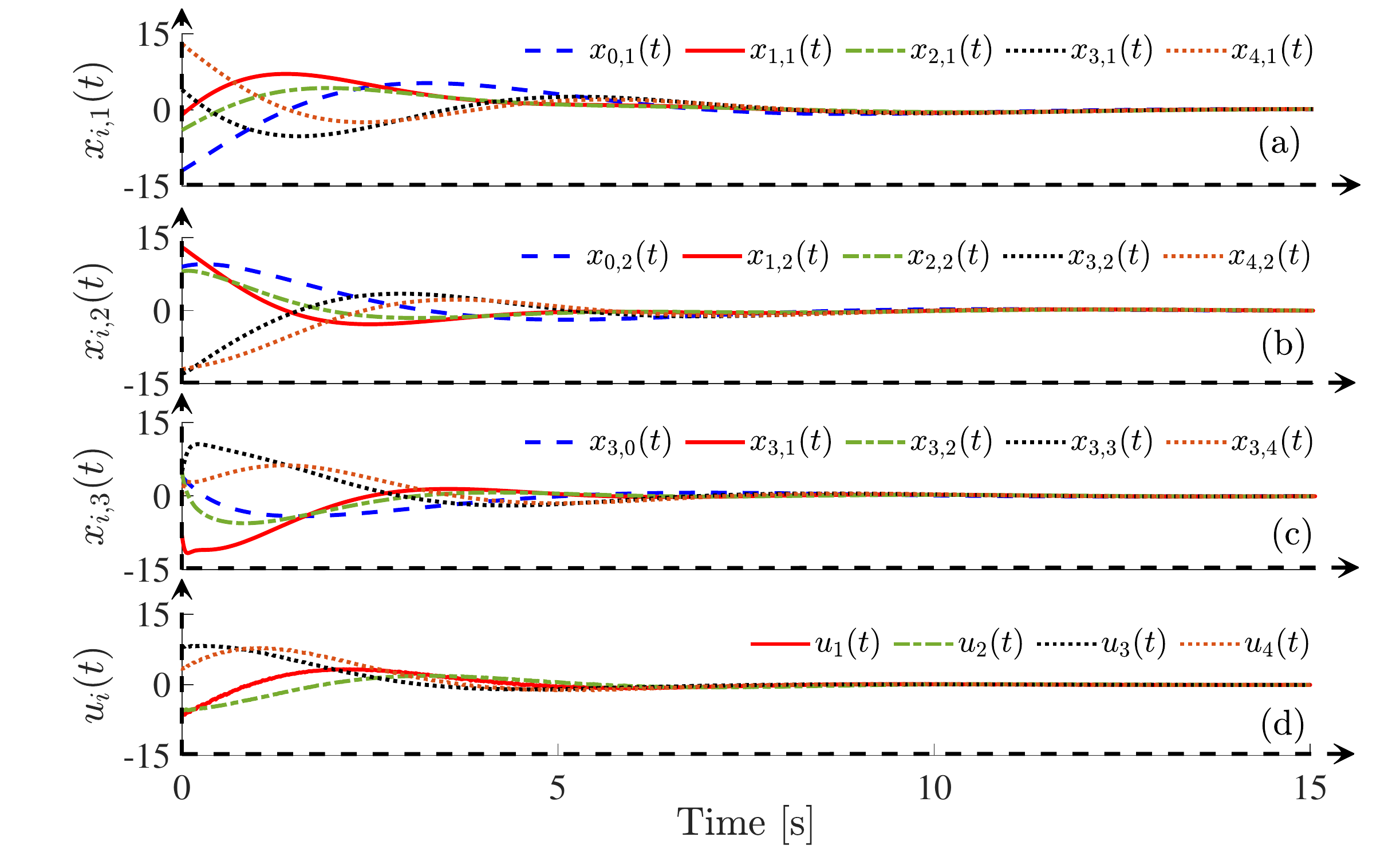}
	\caption{Track of time in (a) the first state variable $x_{i,1}(t)$, (b) the second state variable $x_{i,2}(t)$, (c) the third state variable $x_{i,3}(t)$, and (d) the signal control $u_{i}(t)$. These models are under the linear part of \eqref{eqn:1nn}, and are collected during the communication delays of  $\tau^{\star}=0.8s$ without any perturbations ( $\omega_i(t) = 0$). }
	\label{figurelabe2}
\end{figure}
Numerical calculations were carried out in MATLAB. Tunable gains of the protocol according to the optimization algorithm are obtained  as $k_{10}=\begin{bmatrix}
0.01 &0.011&3.87
\end{bmatrix},~k_{13}=\begin{bmatrix}
0.001  &  0.822 & 0.188
\end{bmatrix},~k_{24}=\begin{bmatrix}
0.01 &   0.01 & 0.143 
\end{bmatrix},~k_{34}=\begin{bmatrix}
0.01 &0.01&0.01
\end{bmatrix},~k_{41}=\begin{bmatrix}
0.01 &0.01&0.01
\end{bmatrix},~k_{42}=\begin{bmatrix}
0.80  &  0.11  &  1.61
\end{bmatrix}$. The initial states equal to $x_{0}=\begin{bmatrix}
-12 &9&4
\end{bmatrix},~x_{1}=\begin{bmatrix}
-1 &13&-8
\end{bmatrix},~x_{2}=\begin{bmatrix}
-4&8&5
\end{bmatrix},~x_{3}=\begin{bmatrix}
4&-13&5
\end{bmatrix}$, and $x_{4}=\begin{bmatrix}
13&-12&0
\end{bmatrix}$, as well as the time derivatives of the communication delays between agents $\tau_{ij}$, that are modeled as random variables with an uniform
discrete distribution in the range of $|\dot{\tau}_{ij}|\leq d_l=\bar{d}_p=1$, so that
conditions 
$\begin{bmatrix}
0 &\tau^{\star}
\end{bmatrix}$ are enforced by means of limiters wherein $\tau^{\star}=0.8s$ obtained by solving Algorithm 1.
{ By randomly selecting the disturbances, defined as biased sinusoidal signals, we have:
\begin{equation}
	\omega_i(t)=a_i+b_i\sin(2\pi f_it) \label{dis}
\end{equation}
within random coefficients of $a_i\in [-3,3]$, $b_i\in [1,6]$ and $f_i\in[1,3]$. The upper bounding constant of the disturbance, $\Gamma$, is obtained from \eqref{dis} as $\Gamma=9$.
\begin{figure}[htp]
	\centering
	\includegraphics[width=1\linewidth,height=0.65\linewidth]{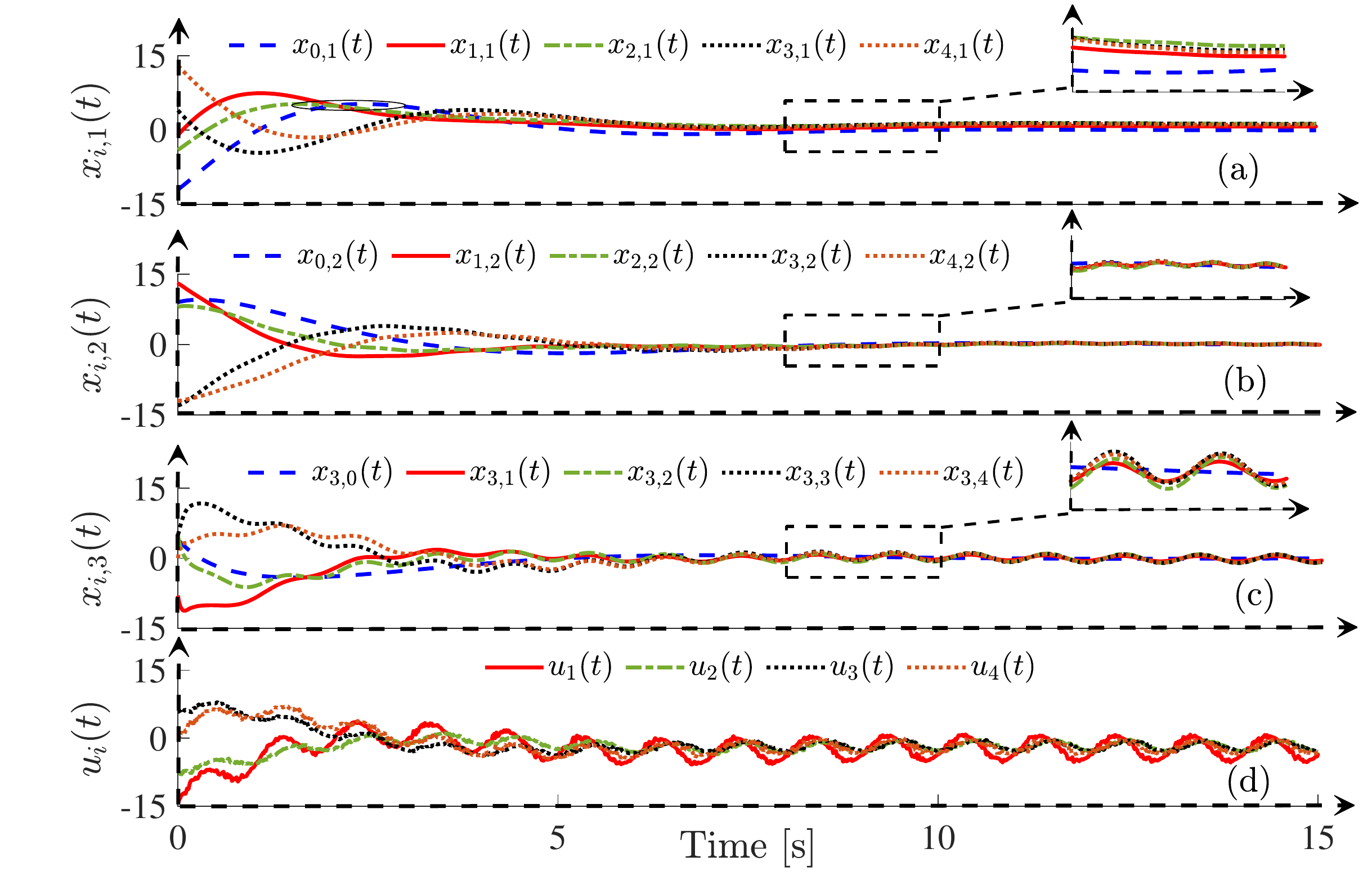}
	\caption{Track of time in (a) the first state variable $x_{i,1}(t)$, (b) the second state variable $x_{i,2}(t)$, (c) the third state variable $x_{i,3}(t)$, and (d) the signal control $u_{i}(t)$. These models are under the linear part of  \eqref{eqn:1nn}, and are collected during the communication delays of $\tau^{\star}=0.8s$ with the perturbations in \eqref{dis}.}
	\label{figurelabe4}
\end{figure}

We should note that the linear part of \eqref{eqn:1nn} is initially tested to acknowledge that the leader’s states are being successfully tracked by the follower’s state variables when the disturbances $\omega_i(t)$ set to zero and $\tau^{\star}=0.8s$.
Time tracking of the follower's state variables compared to the leader is shown in Fig.~\ref{figurelabe2}, supporting the fact that the linear part of the proposed distributed control \eqref{eqn:1nn} is consistent with each of agents tracking the leader's behavior
when the disturbances $\omega_i(t)$ set to zero and $\tau^{\star}=0.8s$. 
Note that the performance of this algorithm is undermined when $\tau^{\star}$ is chosen greater than the optimal value obtained from Algorithm~1 ($\tau^{\star}=0.8s$). As shown in Fig.~\ref{figurelabe4}, the follower’s state variables cannot provide a consensus on the desired
values when the disturbances $\omega_i(t)$ are present.
\begin{figure}[htp]
	\centering
	\includegraphics[width=1\linewidth,height=0.65\linewidth]{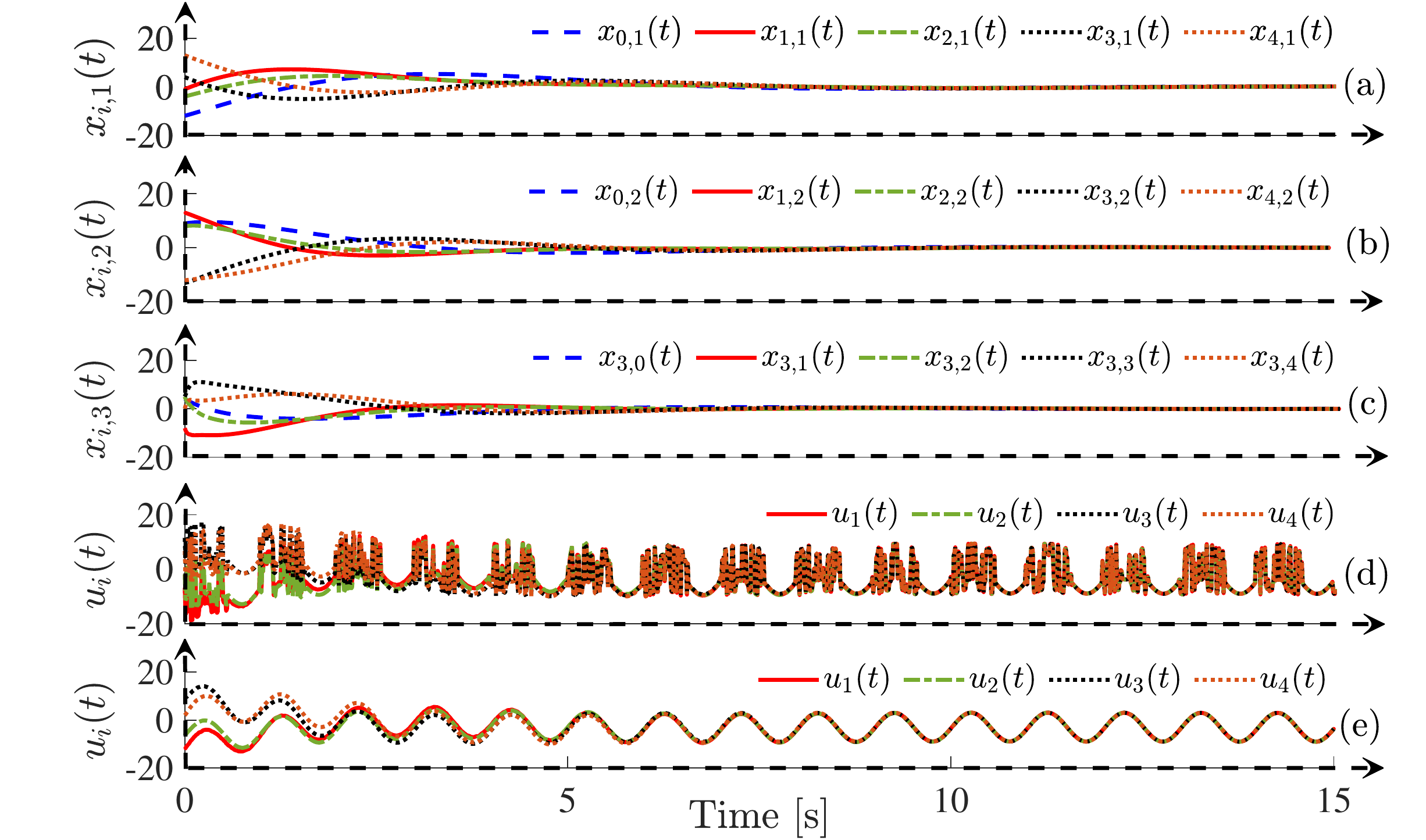}
	\caption{Track of time in (a) the first state variable $x_{i,1}(t)$, (b) the second state variable $x_{i,2}(t)$, (c) the third state variable $x_{i,3}(t)$, (d) the signal control $u_{i}(t)$ with the protocol \eqref{eqn:1nn}, and (e) the signal control $u_{i}(t)$ under the proposed distributed control \eqref{chatring2}, and are collected during the communication delays of $\tau^{\star}=0.8s$ with the perturbations \eqref{dis}. }
	\label{figurelabe6}
\end{figure}

Figs.~\ref{figurelabe6}(a-c) show the agent's states of the distributed control \eqref{eqn:1nn}
upon time-variant delays in a perturbed environment as defined in \eqref{dis}. As expected, the proposed optimization algorithm rejects the perturbations on the agent's dynamics and enables the agent’s systems to track the leader’s states along their communication delays and unknown perturbations. The signal control inputs are therefore not smooth (see Fig.~\ref{figurelabe6}(d)). 
 
To fix chattering problem, the smoothed protocol \eqref{chatring2} is implemented
with the time constant value of $\widetilde{T}=0.01s$ and its performance is verified in during the perturbations.

 The results obtained from Fig.~\ref{figurelabe6}(a-c) and Fig.~\ref{figurelabe6}(e) justify that the proposed distributed control \eqref{chatring2}  correctly
synchronizes the agent’s states to the reference value while fixing the chattering problem along multiple time-variant delays in association with agents and unknown perturbations.
}
\section{Conclusions}
\label{sect7}
This paper investigates a leader-following consensus of MASs in time-variant domains. A new distributed and robust consensus control is jointly programmed under Lyapunov-Krasovskii functional equations and LMIs to show that, with sufficient conditions, the leader-following consensus problem can be solved within multiple time-varying communication delays. The LMI criterion is thus involved to estimate of the upper bound delay in support of the consensus convergence in the system. Furthermore, the optimal constant control gains are mapped to subsequently guarantee the consensus of the MAS to the leader dynamics. These analytical protocols are to broaden their applicability from communication networks to engineered biological systems, where it could lead to the development of new automatic controls in scalable artificial and natural intelligence.
\appendix
\begin{mypf}
	See \cite{gu2003stability} for details.{\hfill $\blacksquare$}
\end{mypf}
\begin{mypflem} See \cite{m1} for details.{\hfill $\blacksquare$}
\end{mypflem}
\begin{mypflem}See \cite{wu2010stability} for details.{\hfill $\blacksquare$}
\end{mypflem}
\begin{mypf}
	Let $X \in \rea^n$ and let $M\in \rea^{n\times n}$ , then
	\begin{equation}\label{e.8}
	M<0 \quad \Leftrightarrow \quad X' M X<0 \quad \forall X\neq 0
	\end{equation}
	According to \eqref{e.8}, we can write
	\begin{equation}\label{e.9}
	X' (M_1+\xi M_2) X= X' M_1 X+ \xi X'  M_2 X<0
	\end{equation}
	By denoting ${\lambda_{max}}^{M_1}<0$ and ${\lambda_{max}}^{M_2}>0$ be the maximum eigenvalues of $M_1$ and $M_2$, \eqref{e.9} can be recast as follows
	\begin{equation}\label{e.10}
	X'M_1 X+\xi  X' M_2 X\leq ({\lambda_{max}}^{M_1}+\xi{\lambda_{max}}^{M_2})\Vert X \Vert^2_2
	\end{equation}
	it follows that
	\begin{flalign}
	{\lambda_{max}}^{M_1}+\xi{\lambda_{max}}^{M_2}&<0 \quad\Rightarrow \quad\xi{\lambda_{max}}^{M_2}<-{\lambda_{max}}^{M_1}\nonumber\\
	\xi&<-\frac{{\lambda_{max}}^{M_1}}{{\lambda_{max}}^{M_2}}=\frac{{\vert\lambda_{max}}^{M_1}\vert}{{\vert\lambda_{max}}^{M_2}\vert}\label{e.11}
	\end{flalign}
	Since $M_1<0,M_2>0$ and $\Vert M \Vert={\lambda_{max}}^{M}$, we can thus rewrite \eqref{e.11} as follows
	\begin{equation}\label{e.12}
	\xi<\frac{\Vert M_1 \Vert}{\Vert M_2 \Vert}
	\end{equation}
	 This concludes the proof.{\hfill $\blacksquare$}
\end{mypf}
\begin{mypf}
{By substituting \eqref{eqn:1nn} into the MAS dynamics \eqref{eqn:1}, we obtain:	
\begin{flalign}
\dot{x}_i(t)=~& \displaystyle Ax_i(t)-B\Big(\sum_{j=0}^{N}\alpha_{ij} k_{ij}
\big( x_{i}(t-\tau_{ij}(t))-x_{j}(t-\tau_{ij}(t))\big)\nonumber\\&+\rho\sign(s_i(t)) -\omega_i(t)\Big) \label{eqn:1newq}
\end{flalign}
and then, by computing the time derivative of \eqref{slide} along with the trajectories of \eqref{eqnnqr}, we reach to:
\begin{equation}
\dot{s}_i(t)=b\Big(\omega_i(t)-\rho\sign\big(s_i(t)\big)\Big)
\end{equation}
Let us now select the following Lyapunov function:
\begin{equation}
\bar{V}(t)\frac{1}{2}\sum_{i=1}^{N} s_i(t)^2
\end{equation}
so that the time derivative of $\bar V(t)$ correspondingly takes the form:
\begin{equation}
\dot {\bar{V}}(t)=\sum_{i=1}^{N} s_i(t)\dot{s}_i(t)=\sum_{i=1}^{N}b\big(s_i(t)\omega_i(t)-\rho|s_i(t)|\big)\label{ew}
\end{equation}
Then by referring to Assumption \ref{ass1},
we manipulate \eqref{ew} as:
\begin{flalign}
\dot{\bar V}(t)\leq -\sum_{i=1}^{N}{b(\rho-\Gamma)}\cdot\left|s_i(t)\right|<0 \hspace{1.5em} \forall~i,\quad
\rho>\Gamma.\label{eqn:301}
\end{flalign}		
so that by reaching to \eqref{eqn:301}, $V(t)=0~\forall~t\geq0$ is concluded. Consequently, the condition $s_i=\dot{s}_i=0$ is invariant since the initial instant of time $t = 0$. Hence, by letting $\dot{s}_i=0$, the following function is supported:
		\begin{flalign}
		\sign(s_i(t))=\frac{1}{\rho} \omega_i(t)\label{neweqn:301}.
		\end{flalign}
Therefore, by substituting \eqref{neweqn:301} into \eqref{eqn:1newq}, we get:}
	\begin{flalign}
	\dot{x}_i(t)=~& \displaystyle Ax_i(t)-B\sum_{j=0}^{N}\alpha_{ij} k_{ij}
	\big( x_{i}(t-\tau_{ij}(t))-x_{j}(t-\tau_{ij}(t))\big)\label{eqn:1newqs}
	\end{flalign}
Let us define errors between the $i$-th and $j$-th agent's states with respect to the leader as
\begin{equation}
	e_{i}(t)=x_{i}(t)-x_{0}\quad,\quad e_{j}(t)= x_{j}(t)-x_{0}
	\label{eqn:n62}
	\end{equation}
after algebraic manipulations, one derives from \eqref{eqn:2} and \eqref{eqn:1newqs} that
\begin{flalign}
	\dot e_i(t)=&Ae_i(t)-B\alpha_{i0}\cdot k_{i0}e_{i}(t-\tau_{i0}(t))\nonumber\\
	&-B\sum_{j=1}^{N}\alpha_{ij}\cdot{k}_{ij}\big((e_{i}(t-\tau_{ij}(t))-e_{j}(t-\tau_{ij}(t))\big)\label{sdot}
	\end{flalign}
Then, by taking \eqref{eqn:66}-\eqref{eqn:n66} into account to, we reaches to:
\begin{align}
	\dot e_i(t)=&Ae_i(t)+A_{i0}e_{i}(t-\tau_{i0}(t))\nonumber\\
	&+\sum_{j=1}^{N}{A}_{ij}\big(e_{i}(t-\tau_{ij}(t))-e_{j}(t-\tau_{ij}(t))\big)
	\label{eqn:67}
\end{align}
To describe the multi-agent dynamics we define the error state vector as
\begin{equation}{\small
		\begin{array}{c}
		e(t)=[\begin{array}{cccc}
		e_{1}(t)' & e_{2}(t)' & \cdots & e_{N}(t)']'
		\end{array}
		\end{array}}\label{eqn:my10}
\end{equation}
Regarding  \eqref{eqn:n51}-\eqref{eqn:n53}, the multi-agent closed loop dynamics can be written as
\begin{equation}
	\dot{{e}}(t)=A_{0} e(t)+\sum_{l=1}^{N}\hat{A}_{l} e(t-\tau_l(t))+\sum_{p=l}^{m}\tilde{A}_{p} e(t-\tau_p(t))
	\label{eqn:68}
\end{equation}
We now present a model transformation. Using the Leibniz Newton formula, it holds:
\begin{flalign}
	e(t-\tau(t))&= {e}(t)-\int_{t-\tau(t)}^{t}\dot{{e}}(s)ds.\label{eqn:75}
\end{flalign}
Thus, the multi-agent closed loop dynamics \eqref{eqn:68} can be transformed to
\begin{flalign}
	{\small\dot e(t)}&={\small Fe(t)-\sum_{l=1}^{N}\hat A_l\int_{t-\tau_l(t)}^{t}\dot e(s)ds-\sum_{p=l}^{m}\hat A_p\int_{t-\sigma_p(t)}^{t}\dot e(s)ds}
	\label{eqn:77}\end{flalign}
where $A, A_{i0}$ and $A_{ij}$ are respectively defined in \eqref{eqn:nn1},\eqref{eqn:66} and \eqref{eqn:n66}. Therefore, to show that $F$ is negative definite, it suffices to prove that $F_{(i,i)}$, for $i = 1,\ldots , N$, is a negative. By considering $k_{ij}>0$ and according to \eqref{eqn:n66} , the term $\sum_{j=1}^{N}{A}_{ij}$ is negative semi-definite.  Consequently, blocks $F_{(i,i)}$ in \eqref{e.fb} are negative definite if the following  matrix $\phi_i= A + A_{i0}$ are negative definite for $i = 1,\ldots, N$.
	\begin{equation}
	\phi_i={\small\begin{bmatrix}
		0 & 1 & 0 &\ldots & 0 \\
		0 & 0 & 1 &\ddots & 0 \\
		\ldots & \ldots & \ldots &\ddots &\vdots\\
		0 & 0 & 0 &\ddots &1\\
		-\gamma_1 & -\gamma_2 & \ldots &\ldots &-\gamma_n
		\end{bmatrix}}\label{phi}\end{equation}
with
\begin{equation}
	\gamma_y=a_y+b\alpha_{i0}\cdot {k}_{i0,y}\quad y=1,\ldots,n.
	\label{phi1} \end{equation}
It is worth mentioning that $F$ is a strictly diagonally dominant block matrix, that its generic block element
$F_{(i,i)} \in \mathbb{R}^{n \times n}$ is defined on the main diagonal as
\begin{equation}
	F_{(i,i)}=A+A_{i0}+\sum_{j=1,i\neq j}^{N}{A}_{ij} \label{e.fb}
	\end{equation}
Let us construct the following Lyapunov-Krasovskii functional
\begin{flalign}
	{V}(e(t))&=  \sum_{i=1}^{5}{V}_{i}(e(t))\label{eqn:78}
\end{flalign}
with
\begin{flalign}
	{V}_{1}(e(t))=&e(t)^{T}P e(t)\label{eqn:79} \\
	{V}_{2}(e(t))=&\sum_{l=1}^{N}\int_{t-\tau_l(t)}^{t}e(s)^{T}Q_le(s)ds\label{eqn:80} \\
	{V}_{3}(e(t))=&\sum_{p=1}^{m}\int_{t-\sigma_p(t)}^{t}e(s)^{T}\bar Q_pe(s)ds\label{eqn:nn80} \\
	{V}_{4}(e(t))=&\sum_{l=1}^{N}\int_{-\tau_l(t)}^{0}\int_{t+\theta}^{t}\dot{e}(s)'R_l\dot{e}(s)dsd\theta\label{eqn:82}\\
	{V}_{5}(e(t))=&\sum_{p=1}^{m}\int_{-\sigma_p(t)}^{0}\int_{t+\theta}^{t}\dot{e}(s)'\bar R_p\dot{e}(s)dsd\theta\label{eqn:nn82}
\end{flalign}
where, in accordance with the statement of Theorem~\ref{thm2}, $P,Q_l,\bar Q_p,\bar R_p$ and $R_l\in \mathbb{R}^{Nn\times Nn}$ are symmetric positive definite matrices. Now, the time derivative of ${V}_{1}( e(t))$ in \eqref{eqn:79} along the trajectories of the system in \eqref{eqn:77} are given by
\begin{align}
	\dot{{V}}_{1}(e(t))=&e(t)'(F'P+PF)e(t)-2e(t)'P\sum_{l=1}^{N}\hat A_l\int_{t-\tau_l(t)}^{t}\dot e(s)ds\nonumber\\&-2e(t)'P\sum_{p=1}^{m}\tilde A_p\int_{t-\sigma_p(t)}^{t}\dot e(s)ds\label{eqn:83}
\end{align}
According to \eqref{Eeq:2lem2} in Lemma~\ref{lem2}, \eqref{eqn:83} can be rewritten as
\begin{flalign}
	\dot{{V}}_{1}(e(t))\leq~& e(t)'( F'P+PF) e(t)+e(t)'( \sum_{l=1}^{N}\tau^{\star} P\hat A_lR_l^{-1}\hat A_l'P) e(t)
	\nonumber\\&+\sum_{l=1}^{N}\int_{t-\tau_l(t)}^{t}\dot{e}(s)R_l \dot{e}(s)ds+\sum_{l=1}^{N}\int_{t-\tau_l(t)}^{t}\dot{e}(s)R_l \dot{e}(s)ds	\nonumber\\&+e(t)'\Big(\sum_{p=1}^{m}\tau^{\star} P\tilde A_p \bar R_p^{-1}\tilde A_p'P\Big) e(t)
	.\label{eqn:84}
\end{flalign}
From \eqref{eqn:80} and \eqref{eqn:nn80}, by differentiating ${V}_{2}(e(t))$,${V}_{3}(e(t))$ and exploiting the bound on delays according to Assumption~\ref{ass2}, we get
\begin{equation}
	\dot {{V}}_{2}(t)\small{\leq  e(t)'\sum_{l=1}^{N}Q_le(t)-\sum_{l=1}^{N}e(t-\tau_l(t))'Q_l(1-d_l)e(t-\tau_l(t))}\label{eqn:85}
\end{equation}
\begin{equation}
	\dot {{V}}_{3}(t)\small{\leq e(t)'\sum_{p=1}^{m}\bar Q_pe(t)-\sum_{p=1}^{m}e(t-\sigma_p(t))'\bar Q_p(1-\bar d_p)e(t-\sigma_p(t))}\label{eqn:8nn5}
\end{equation}
By taking the time derivative of ${V}_{4}(e(t))$ and ${V}_{5}(e(t))$, it yields to
\begin{flalign}
	\dot {{V}}_{4}(t)&= \sum_{l=1}^{N}\tau_l(t)\dot{e}(t)'R_l\dot{e}(t)- \sum_{l=1}^{N}\int_{t-\tau_l(t)}^{t}\dot{e}(s)'R_l \dot{e}(s)ds\label{eqn:86}
	\\
	\dot {{V}}_{5}(t)&=\sum_{p=1}^{m}\sigma_p(t)\dot{e}(t)'\bar R_p\dot{e}(t)-\sum_{p=1}^{m}\int_{t-\sigma_p(t)}^{t}\dot{e}(s)'\bar R_p \dot{e}(s)ds\label{eqn:nn86}
\end{flalign}
Considering the upper bound of time-delays $\tau^{\star}$, we can write
\begin{flalign}
	\dot {{V}}_{4}(t)&\leq \sum_{l=1}^{N}\tau^{\star}\dot{e}(t)'R_l\dot{e}(t)- \sum_{l=1}^{N}\int_{t-\tau_l(t)}^{t}\dot{e}(s)'R_l \dot{e}(s)ds\label{eqn:NE86}\\
	\dot {{V}}_{5}(t) &\leq\sum_{p=1}^{m}\tau^{\star}\dot{e}(t)'\bar R_p\dot{e}(t)- \sum_{p=1}^{m}\int_{t-\sigma_p(t)}^{t}\dot{e}(s)'\bar R_p \dot{e}(s)ds\label{eqn:NEWnn86}
\end{flalign}
Next, by summing \eqref{eqn:84}-\eqref{eqn:NEWnn86} and from \eqref{H}-\eqref{m1T}, one derives	
\begin{flalign}
	\dot V(e(t))\leq&  e(t)' \Big(\sum_{l=1}^{N}\tau^{\star} P\hat A_lR_l^{-1}\hat A_l'P+\sum_{p=1}^{m}\tau^{\star} P\tilde A_p \bar R_p^{-1}\tilde A_p'P+M_1 \Big)e(t)\nonumber\\
	&+\tau^{\star}\dot{e}(t)'H\dot{e}(t)+e(t-\tau_l(t))'\bar M_1e(t-\tau_l(t))\nonumber\\
	&+e(t-\sigma_p(t))'\tilde M_1e(t-\sigma_p(t))\label{eqn:88}\end{flalign}
Let us expand the term $\tau^{\star}\dot{e}(t)'H\dot{e}(t)$ according to \eqref{eqn:68} as follows
	\begin{flalign}
	\tau^{\star}\dot{e}(t)'H\dot{e}(t)=~&\tau^{\star}{e}(t)'\Big(A_0'HA_0\Big){e}(t)\nonumber\\&+\tau^{\star}\sum_{l=1}^{N}{e}\big(t-\tau_l(t)\big)'\Big(\hat A_l'H\hat A_l\Big){e}\big(t-\tau_l(t)\big)\nonumber\\
	&+\tau^{\star}\sum_{p=1}^{m}{e}\big(t-\sigma_p(t)\big)'\Big(\tilde A_p'H\tilde A_p\Big){e}\big(t-\sigma_p(t)\big)\nonumber\\
	&+2\tau^{\star}{e}(t)'\Big(A_0H\sum_{l=1}^{N}\hat A_l\Big){e}\big(t-\tau_l(t)\big)\nonumber\\
	&+2\tau^{\star}{e}(t-\tau_l(t))'\Big(\sum_{l=1}^{N}\hat A_l'H\sum_{p=1}^{m}\tilde A_p\Big){e}\big(t-\sigma_p(t)\big)\nonumber\\&+2\tau^{\star}{e}(t)'\Big(A_0H\sum_{p=1}^{m}\tilde A_p\Big){e}\big(t-\sigma_p(t)\big)\label{eq:m1}
	\end{flalign}
From \eqref{Eeq:1lem2} (see Lemma~\ref{lem2}), and assuming
	\begin{flalign}
	a(t)&=A_0e(t),\quad b(t)=H\sum_{l=1}^{N}\hat A_l{e}(t-\tau_l(t)),\quad
	\psi= H, \nonumber \\\bar a(t)&=A_0e(t),\quad 
	\bar b(t)=H\sum_{p=1}^{m}\tilde A_p{e}(t-\tau_p(t)),\quad\nonumber\\
	\hat a(t)&=\sum_{l=1}^{N}\hat A_le(t-\tau_l(t)),\quad \hat b(t)=H\sum_{p=1}^{m}\tilde A_p{e}(t-\tau_p(t)).
\end{flalign}
the following results are then obtained	
	\begin{flalign}
	2a(t)'b(t)\leq~& e(t)'A_{0}HA_{0}e(t)\nonumber\\&+ e\big(t-\tau_l(t)\big)'\Big(\sum_{l=1}^{N}\hat A_l'H\hat A_l\Big)e\big(t-\tau_l(t)\big)\label{eq:nm}\\
	2\bar a(t)'\bar b(t)\leq~& e(t)'A_{0}HA_{0}e(t)\nonumber\\&+ e\big(t-\sigma_p(t)\big)'\Big(\sum_{p=1}^{m}\tilde A_p'H A_p\Big)e\big(t-\sigma_p(t)\big)\label{eq:nm1}\\
	2\hat a(t)'\hat b(t)\leq~& e(t-\tau_l(t))'\Big(\sum_{l=1}^{N}\hat A_l'HA_{l}\Big)e(t-\tau_l(t))\nonumber\\&+ e\big(t-\sigma_p(t)\big)'\Big(\sum_{p=1}^{m}\tilde A_p'H\tilde A_p\Big)e\big(t-\sigma_p(t)\big)\label{eq:nm2}
	\end{flalign}
Therefore, according to \eqref{eq:nm}-\eqref{eq:nm2}, \eqref{eq:m1} can be recast as
	\begin{flalign}
	3\tau^{\star}\dot{e}(t)'H\dot{e}(t)\leq~&\tau^{\star}{e}(t)'\Big(A_0'HA_0\Big){e}(t)\nonumber\\
	&+3\tau^{\star}{e}(t-\tau_l(t))'\Big(\sum_{l=1}^{N}\hat A_l'H\hat A_l\Big){e}(t-\tau_l(t))\nonumber\\
	&+3\tau^{\star}{e}(t-\sigma_p(t))'\Big(\sum_{p=1}^{m}\tilde A_p'H\tilde A_p\Big){e}(t-\sigma_p(t))\label{eq:m2}
	\end{flalign}	
Now, by substituting \eqref{eq:m2} into \eqref{eqn:88}, one derives
	\begin{flalign}
	\dot V(e(t))\leq~&e(t)'H_1e(t)
	+e(t-\tau_l(t))'H_2e\big(t-\tau_l(t)\big)\nonumber\\&+e\big(t-\tau_p(t)\big)'H_3e\big(t-\tau_p(t)\big)\label{eqn:N88}
	\end{flalign}
	being
	\begin{flalign}
	H_1=M_1+\tau^{\star}M_2, ~ H_2=&\bar M_1+\tau^{\star}\bar M_2,~H_3=\hat M_1+\tau^{\star}\hat M_2\label{eqn:89}
	\end{flalign}
	with
	\begin{flalign}
	M_2&=3A_{0}'HA_{0}+\sum_{l=1}^{N} P\hat A_lR_l^{-1}\hat A_l'P+\sum_{p=1}^{m}P\tilde A_p \bar R_p^{-1}\tilde A_p'P \nonumber\\
	\bar M_2&=3\sum_{l=1}^{N}\hat A_l'H\sum_{l=1}^{N}\hat A_l,\quad \quad
	\tilde M_2=3\sum_{p=1}^{m}\tilde A_p'H\sum_{p=1}^{m}\tilde A_p\nonumber
	\end{flalign}
Therefore, to have $\dot V(e(t))\leq 0$, $H_1, H_2$ and $H_3$ in \eqref{eqn:N88} are to be negative definite. It should be noted that  $H_1$ is a non-linear inequality due to the presence of terms $R_l^{-1}$ and $\bar R_p^{-1}$. Therefore, performing the Schur complement on $H_1$ (see Lemma~\ref{lem1}), \eqref{eqn:N88} can be rewritten as in \eqref{eqn:49}-\eqref{eqn:50}, which consists of LMIs. Their solutions can be easily found by using standard numerical solvers based on the the interior point method. Hence, if \eqref{eqn:49}-\eqref{eqn:50} are satisfied, then $\dot V(e(t))\leq 0$ in \eqref{eqn:N88} and it results $  {V}(e(t))$ converges to zero and thus condition \eqref{eqn:syn} is in force. This concludes the proof.{\hfill $\blacksquare$}
\end{mypf}
\section*{Acknowledgment}
We thank Alessandro Pilloni for the useful discussions on developing the optimization algorithm.
\ifCLASSOPTIONcaptionsoff
  \newpage
\fi

\bibliographystyle{IEEEtran}
\bibliography{biblio}

\vfill
\enlargethispage{2in}
\end{document}